# Tunable colloidal photonic crystals

Talha Erdem[1,‡], Mykolas Zupkauskas[1, ‡], Thomas O'Neill[1, ‡], Alessio Caciagli[1, ‡], Peicheng Xu[1], Yang Lan[2], Peter Boesecke[3] and Erika Eiser[1,‡*]

[1]Cavendish Laboratory, Department of Physics, University of Cambridge, JJ Thomson Avenue, Cambridge CB3 0HE, United Kingdom

[2]Department of Chemical and Biomolecular Engineering, Towne Philadelphia, PA 19104, University of Pennsylvania, U. S. A.

[3]European Synchrotron Radiation Facility, avenue des Martyrs 71, 38043 Grenoble Cedex 9, France

[‡]These authors contributed equally.

*Corresponding Author: ee247@cam.ac.uk

Spherical colloids arranged in a crystalline order are known to produce structural colors. The intensity and brilliance of such photonic crystals require high size-monodispersity of the colloids, a low number of lattice defects and disorder, as well as a relatively large refractive index contrast between the scattering colloids and the continuous background. Here we present the unexpected photonic properties of aqueous suspensions of charge-stabilized, 186 nm large, fluorinated colloids with a refractive index of 1.37. Employing reflectivity, optical observation, small angle x-ray scattering measurements and reflectivity modeling, we demonstrate that these suspensions become partially transparent while showing strong, almost angle-independent color in reflection despite the very small refractive index difference. Under certain conditions additional sharp Bragg reflections are observed. We were able to tune the observed structural colors continuously across the entire visible range by simply changing the volume fraction of these colloidal suspensions, which show a white appearance when dilute, structural color and Bragg peaks when concentrated enough to form Wiegner crystal, and angle-independent color when very concentrated and in a glassy phase.

**1. INTRODUCTION**

Colloidal particles are omnipresent in our daily live, being a major ingredient in cosmetics,[1] agriculture,[2] bioimaging,[3] and electronics,[4] to mention a few. Due to their size ranging from tens of nanometer to micrometer they are also known to form structural colors when forming photonic crystals,[3,5–8] and therefore inspired intensive research in colloid cystallization.[9–15] Similar to natural colloidal crystals, their artificial counterparts made of a variety of materials such as inorganic nanocrystals,[16,17] biological materials,[18,19] and polymers[20–22] have also attracted significant interest. Many of these studies were instrumental in the understanding of nucleation and growth phenomena and the design of systems with interesting optical, electronic, or mechanical properties.[8,23–25]

Hard, spherically symmetric colloids that are larger than 100 nm can form crystals with photonic activity in the visible range with either face-centered cubic (FCC), hexagonal close-packed (HCP), or random hexagonal closed-packed (RHCP) structures.[26–28] On the other hand, softer spheres with weak, long ranged repulsive Coulomb interactions can form body-centered cubic (BCC) and other lattice structures, depending on the strength of the interaction.[29–31] Using long polymers grafted to nanoparticles[32] is another means to introduce 'soft' interactions.

The type of lattice formed as well as the size and refractive index of the colloids govern the optical features of these photonic systems. For instance, the close-packed FCC lattices made of a single type of colloid cannot open a full bandgap in the visible range regardless of the refractive index difference between the medium and the colloids.[33] Nevertheless, strong reflections can be observed at certain angles.[34] In addition to the lattice type, the refractive index contrast, $\Delta n$, between the colloid and the surrounding medium is an essential parameter for designing materials with structural color. Finlayson et al.[2] demonstrated that while the strength of the reflection in photonic crystals is expected to have a $(\Delta n)^2$ dependence due to Fresnel reflections, the distortions in the lattice structure and the size distribution of the colloids change this dependence to a linear one.[35] The authors also showed that when $\Delta n = 0.05$, the reflectance from the solid colloidal photonic crystals is <10%; this is accompanied by a strong angle dependence.[36] These results show that satisfying the transmission and reflection over a broad angular range at the same time is challenging.

Here we demonstrate that these challenges can be overcome by using aqueous suspensions of Fluorinated Latex (FL) particles with low $\Delta n$. We show that these charged colloids form almost transparent Wigner crystals and glassy structures in water with tunable interparticle distances. These ordered structures appear blue, green and red in reflection depending on the colloidal volume fraction. These weakly angle-depended structural colors are accompanied by intense Bragg reflections under certain circumstances. Using electromagnetic simulations we can confirm that the 3D photonic crystals formed are capable of strongly reflecting light at certain wavelengths despite the very low refractive index difference. We believe that the work presented here will be of interest in developing new displays, wearable electronics, and optical coatings.

## 2. EXPERIMENTAL SECTION

**2.1 Synthesis of Fluorinated Latex and Polystyrene Particles.** The FL particles were synthesized using emulsion polymerization detailed in ref. [37]: 2 mg of the surfactant sodium dodecyl sulphate (SDS) was dissolved in 12 mL of deionized (di-)water. Next, 325 mg of the colloid forming 2,2,3,3,4,4,4-Heptafluorobutyl methacrylate (HFBMA; obtained from Alfa Aesar) was added and emulsified at 70 °C in nitrogen atmosphere while stirring at 600 rpm for 1 h. 7.75 mg of the initiator and stabilizer potassium persulfate (KPS) per 500 μL di-water solution was then injected into the HFBMA and SDS mixture while stirring at reduced speed. The reaction continued for 18 h at 70 °C. The synthesized particles were purified by dialysis and then stored in a refrigerator. SDS and KPS were purchased from Sigma-Aldrich.

Negatively charged polystyrene (PS) particles used here were purchased from Cambridge Bespoke Colloids LCC.

**2.2 Scanning and Transmission Electron Microscopy Imaging.** The FL particles were imaged using a Zeiss scanning electron microscope (SIGMA VP field emission SEM), while the PS particles were imaged with a transmission electron microscope (TEM, TECNAI 20 from FEI of Thermo Fisher Scientific).

**2.3 DLS and Zeta Potential Measurements.** We used a Malvern ZetaSizer Nano ZSP to measure the particle size and size-distribution. In order to determine the Zeta potential we dispersed the particles in 10 mM phosphate buffer, while all other results were obtained from suspensions in pure di-water.

**2.4 Reflectivity Measurements.** The reflection measurements were taken using a reflection probe bundle equipped with different LED light sources and a spectrometer. Illumination and collection was done at normal angle. The green and red angles were illuminated with a fiber-coupled white LED (Thorlabs MCWHF2) and the blue sample was illuminated with a cyan LED emitting at 490 nm (Thorlabs M490F3). The spectra were measured using a Thorlabs CCS 100 spectrometer. Reference measurements were taken using a Thorlabs BB1-E02 dielectric mirror.

**2.5 Small-Angle X-ray Scattering.** SAXS measurements were carried out at the ID2 Beamline of the European Synchrotron Radiation Facility (ESRF) in Grenoble, France and at the B21 station of the Diamond light source, UK. The 2D spectra measured at the ESRF were obtained using X-rays with a wavelength of 0.1 nm and a beam size of 50 μm × 20μm on the sample. The sample to detector distance was 10 m and the camera used was a Rayonix MX170 detector. The setup at Diamond used the same wavelength X-rays but a beam size of 1 mm × 1 mm, a sample to detector distance of 4 m and a Pilatus 300 K detector. All spectra were normalized against the background and empty cell.

**2.6 Electromagnetic simulations.** Electromagnetic simulations were carried out using a commercially available finite-different time-difference simulator in three dimensions (Lumerical Inc.). Here, latex particles were modeled as dielectric spheres with a refractive index of 1.37 while the background index was set to 1.33 to mimic water. The colloidal lattice constant was varied between 300 and 500 nm. The structure was illuminated at normal angle by a broadband plane wave source. Bloch boundary conditions were applied at the *x*- and *y*-boundaries of the simulation region while the *z*-boundary was set as a perfectly matched layer.

## 3. RESULTS AND DISCUSSION

**3.1 Particle Synthesis and Characterization.** The FL particles were synthesized using an emulsion polymerization technique.[37] The reaction relies on the polymerization of a fluorinated acrylate (HFBMA) in the presence of the anionic surfactant SDS acting as stabilizer, and KPS being the initiator. The obtained FL particles have a refractive index of 1.37,[38] which is very close to the refractive index of water (1.33 at 20°C). We measured the zeta potential of our colloids as (-82 ± 4) mV, while DLS measurements provided us with an average hydrodynamic diameter of 193 nm with a polydispersity index of 1.6% (Fig. 1a,b). Note that SEM images of the dried, densely packed colloidal samples suggested an average particle size of (160 ± 25) nm and a particle-to-particle distance of 166±0.7nm (inset in Fig. 1a). The Form Factor, $F(q)$, obtained for dilute suspensions with a colloidal volume fraction of $\phi \approx 0.1\%$ shed light on this size difference. Here, $q$ is the scattering vector. Analyzing the scattered intensity obtained in SAXS measurements revealed a bimodal distribution with 70% of the signal accounting for FL particles with a diameter of 183 nm while the remaining particles have a rather narrow size distribution around 134 nm. Details are given in the section on SAXS. Note that dried films of FL particles appear completely white at all angles except in a very narrow angular range where a strong purple Bragg reflection showed. In Fig. 1 we also present the Zeta potential (-30 ± 4) mV and hydrodynamic size (220nm; polydispersity index 1.4%) measurements for the PS particles used. TEM images give a dry particle size of (190 ± 5) nm suggesting that the crosslinked PS particles 'shrink' less strongly than the FL ones when dried. Hence in aqueous solution the two particle sizes are similar but still more than 10% different. Dry films of these PS-particles show an even stronger white appearance with a sharp Bragg reflection in the blue resulting from the ordering of the colloids into hexagonally close-packed layers at the free film-air interface. The white background is due to the strong scattering from the PS particles, which have a refractive index of 1.56, and disorder in form of grain boundaries and lattice defects generated during the drying process.

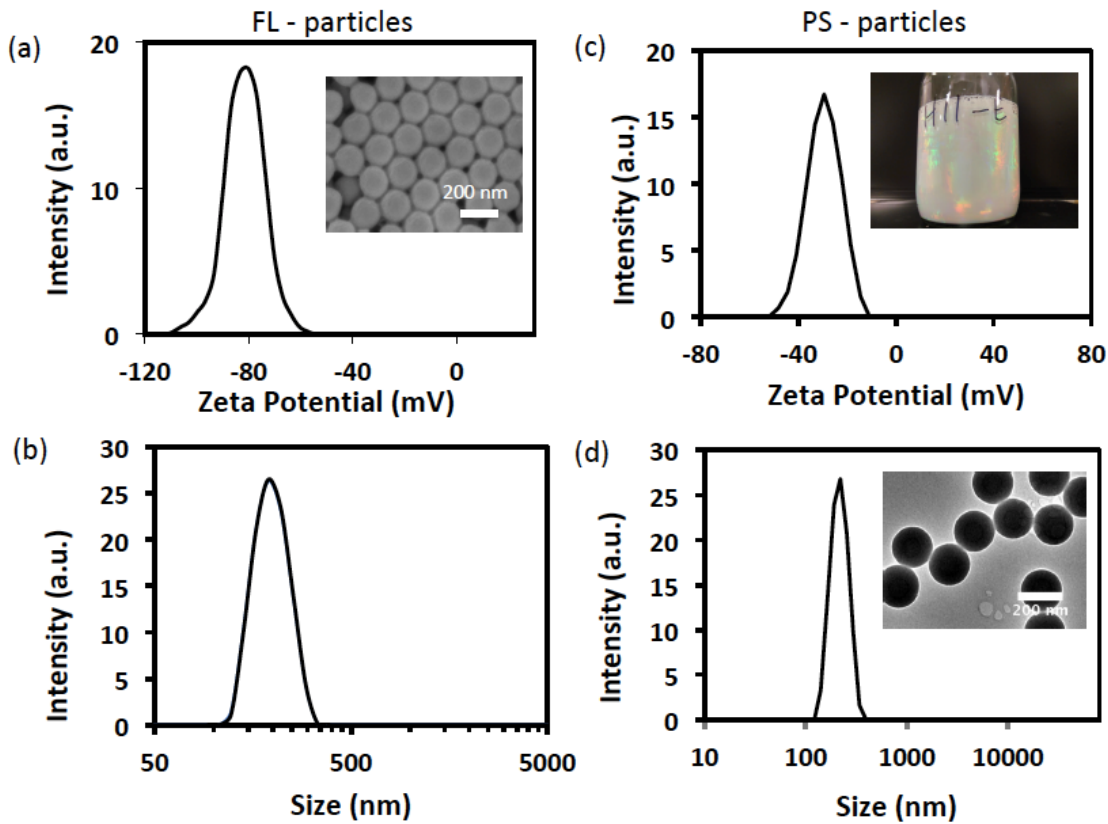

**Figure 1.** (a) Zeta-potential measurement of the synthesized FL particles suspended in 10 mM phosphate buffer. The inset shows an SEM image of the dried particles. (b) Size distribution of the FL particles measured using DLS. (c) Zeta-potential measurement of our PS particles suspended in 10 mM phosphate buffer. The inset shows a photograph of these PS particles after synthesis and washing. The particle content is roughly 2%, but still the strongly scattering dispersion shows iridescent colors. (d) Size distribution of the PS particles measured using DLS. The inset shows a TEM micrograph taken of the dried PS particles.

**3.2 Reflectivity and Transmission of FL and FL+PS Suspensions.** Both dilute, aqueous suspensions of FL and PS particles appear completely white. Nevertheless, both dispersions already showed some structural color in dilute solutions with only about 2% colloidal volume fractions due to their strong charging state and their ordering close to the container walls (see inset in Fig. 1c). Upon concentrating the solutions by centrifugation, the FL suspensions became significantly more transparent and overall colored while both, pure PS and PS+FL suspensions remained predominantly white and non-transparent. For better visualization and for the SAXS measurements shown later we filled the concentrated samples into flat 200 μm thick, 4 mm wide and 30 mm long glass capillaries, which were then sealed with a two-component glue. In Fig. 2a the photographs of FL samples displayed a strong bluish reflection for an estimated ~40%, greenish at volume fractions of around 26-33% and red for 20-23%. The transmission measurements of these particles that are measured using UV-Vis spectrophotometry over a path length of 1 cm show that the transmission decreased with decreasing concentration in the range of 40-20% (Fig. 2b and Table 1). Below around 15%, the FL suspensions turned white and non-transparent again, suggesting a transition from an ordered to a more liquid, disordered state. For comparison, we investigated suspensions of PS particles of very similar size and Zeta potential. As expected, the PS suspensions with a $\Delta n_{W\text{-}PS} \approx 0.23$ remained predominantly white at concentrations between 2% and about 40% with a strongly angle-dependent color due to Bragg diffraction again in the range of 20-40%. Further we show in Fig. 2 the transmission measurements of 1:1 mixture of FL and PS particles testing whether reducing the suspensions' refractive index difference $\Delta n$ would lead to some transparency and structural color at volume fractions between 20% and 40%. Although also these mixtures where white at most angles and non-transparent they did show some strong Bragg reflections within a very narrow angular range, which we will discuss in the SAXS section.

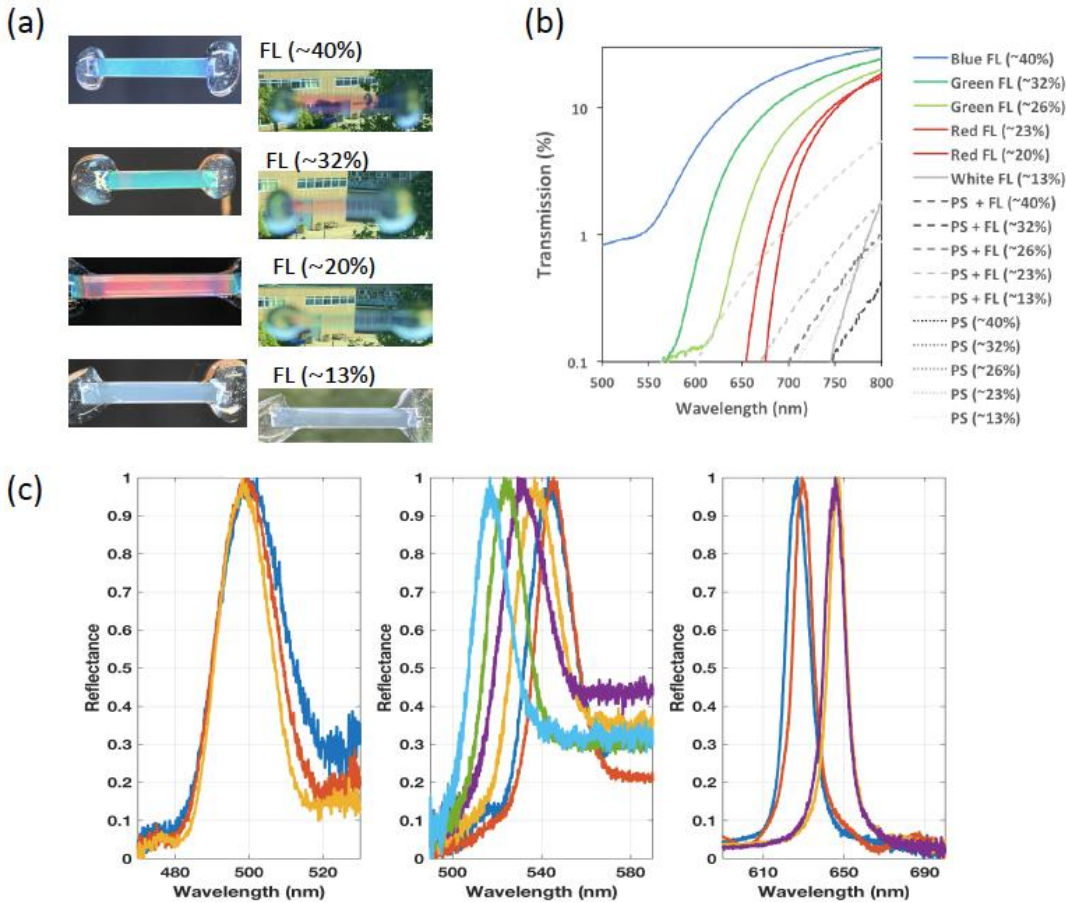

**Figure 2.** (a) Photographs of flat cuvettes containing differently concentrated, aqueous suspensions of about 186 nm large FL particles, taken in reflection (right) and transmission (left). (b) Transmission spectra measured using a UV-Vis spectrophotometer over a path length of 1 cm containing different volume fractions of FL, FL+PS and only PS particles. (c) Normalized reflectivities measured at different locations in the (left) blue, (middle) green and (right) red sample of the FL suspensions shown in (a).

Compared to previous reports on colloidal crystals that typically show a white background with sharp Bragg reflection, our FL suspensions display three interesting properties: Enhanced transparency for volume fractions ranging between 15% and 40%. Secondly, when looking at the sample with the naked eye in reflection, we observe a relatively angle-independent color. Thirdly, some samples show an additional strong Bragg reflection within a very narrow angular range typical for most photonic crystals made of spherical particles, but those that did not show such Bragg reflection still appeared colored and partially transparent. Interestingly, the angle-independent color seen in reflection by eye showed a small variation in the position of the reflection peaks, depending on the location we measure in the sample. Such a weak dependence suggests the existence of multiple crystalline domains having similar but slightly different lattice constants that most likely originates from both fractionation due to small variations of surface charges from colloid to colloid and the bimodal size distribution of our FL particles. Having multiple crystal domains of similar characters enables single color reflection at different angles rather than iridescence.

**Table 1.** Measured reflection peaks and line widths.

|  | Peak (nm) | Line width (nm) | Line width (meV) |
|---|---|---|---|
| Blue | 498 | 16.6 | 83 |
|  | 499.4 | 19.5 | 97 |
|  | 500.2 | 22.4 | 111 |
| Green | 543.1 | 25.2 | 105 |
|  | 545.3 | 21.2 | 88 |
|  | 536.9 | 27.1 | 116 |
|  | 531.5 | 30.4 | 132 |
|  | 524.8 | 22.8 | 102 |
|  | 516.8 | 22.6 | 104 |
| Red | 627 | 13.6 | 43 |
|  | 629.1 | 11.5 | 36 |
|  | 647.2 | 10.5 | 31 |
|  | 645.6 | 12.0 | 36 |

**3.3 Small-Angle X-ray Scattering.** Following the reflectivity measurements that indicate that the charge-stabilized FL colloids form ordered structures we performed SAXS measurements using the focused beam of the ID2 beamline at the ESRF. In Figure 3 we show the two-dimensional SAXS patterns and the corresponding structure factors, $S(q)$, that were obtained by azimuthally integrating the scattering intensities, $I(q)$, and dividing these with the separately measured form factor, $F(q)$. Suspensions with a volume fraction of about 40% containing only FL particles and a 1:1 mixture of similarly sized PS and FL particles are shown. As expected, the pure FL suspension displayed very clear Bragg peaks, arising from the (111), (220), (113), and (133) scattering planes of an FCC lattice corresponding to scattering vectors $q_0 = 0.0403$ nm$^{-1}$, $q_1 = 0.0657$ nm$^{-1}$, $q_2 = 0.0765$ nm$^{-1}$ and $q_3 = 0.1013$ nm$^{-1}$. The strong hexagonal arrangement of the peaks stems from the dense (111) planes aligning parallel to the confining flat capillary walls. One could be tempted to think that this strong scattering came from a single crystal. Inspecting both the 2D spectrum and the radial integration more closely, we make three observations. First the Bragg peaks are slightly spread radially, which indicates a slight spread of the rotational orientation of the (111) planes. Hence it would be surprising to have a single crystal in the sample considering the sample thickness is 200 μm and the beam size to be 150×150 μm. Note that similar scattering peaks with identical position were obtained in several spots measured in the sample. Secondly, there are also faint hexagonal scattering patterns like rotated shadows of the strong peaks but at slightly higher $q$ values. This may be due to the formation of a smaller crystal but with slightly larger lattice spacing and orientation. The difference in lattice spacing can be understood in terms of the bimodal size distribution across all FL colloids, leading to local fractionation. Further, we initially estimated the sample volume fraction to be roughly 30%, but we compute from the main peaks at $q_0 = 0.0403$ nm$^{-1}$ the separation between the centers of two neighboring colloids in the FCC lattice to be 191nm corresponding to 50%. This also means two colloid surfaces are some 30 nm apart. Note that such a big variation between estimated and measured volume fraction is due to the fact that dense colloidal suspension evaporates much more rapidly during the filling of the capillaries than dilute ones.

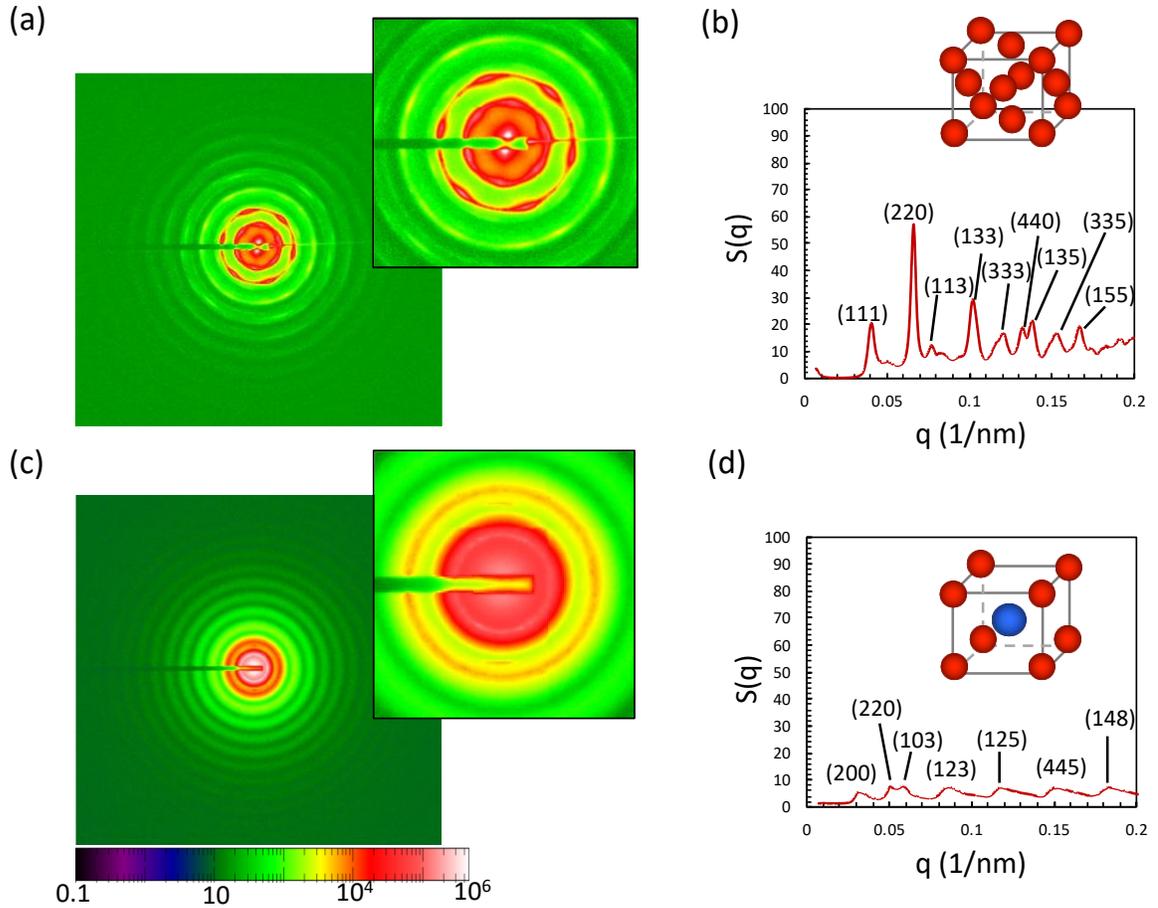

**Figure 3.** (a) SAXS spectrum taken of a 40% FL particle suspension. (b) Azimuthally integrated structure factor of this sample, after dividing out the form factor, measured in a 0.1% suspension. The peaks are assigned to the corresponding Miller indices (*hkl*). (c) Corresponding SAXS and (d) integrated spectrum of the 1:1 mixture made of FL and PS particles.

Although it is known that FCC crystals of hard spheres never form for size polydispersities larger than 6%, it will not influence the crystal formation in a Wiegner crystal.[39] The difference in lattice spacing observed in our system is rather due to the fact that the colloids will have a charge distribution. As the suspension of our charged FL colloids has the possibility to rearrange we observe local fractionation into local crystals of similarly sized particles. Thirdly, the *S(q)* spectra show many more peaks at larger *q*'s, which we could not assign to the FCC structure but rather to a BCC symmetry very similar to the situation observed by Sirota et al.,[31] who measured at higher volume fractions and low ionic strength a coexistence between FCC and BCC crystals and a glassy texture. In particular latter is typical for metallic glasses made of a single component. This coexistence of different crystal structures and lattice variation within a crystal structure due to size or charge polydispersity in combination of a small refractive index variation explains why we observe under certain angles strong iridescent colors while the samples show on the whole rather angle-independent structural colors and significant transparency.

Indeed, Sirota et al. performed systematic SAXS measurements of charged PS-particle suspensions, establishing a phase diagram by plotting the measured structure as function of the ionic strength in the sample.[31] At very low ionic strengths (deionized solution) they observe a first transition from a liquid, disordered phase to a BCC crystal followed by a coexistence region between BCC and FCC structure as the colloid volume fraction is increased from <5% to 6-15% and 16-22% respectively. At even higher volume fractions the dispersions became glassy, manifesting the glassy character in terms of retaining some of the FCC-crystal peaks that started to broaden. When increasing the ionic strength to about 0.2 mM the BCC phase disappears completely and the transitions from a liquid to FFC and subsequently to the glassy phase shifted to higher volume fractions. Increasing the added salt concentration slightly can be also interpreted in terms of making the repulsive Coulomb interactions more hard-sphere like, as the Debye screening length becomes more short-ranged but steeper. With a zeta potential of around -80 mV our FL particles show a very similar behavior in dialyzed aqueous suspensions, although at shifted volume fractions. Like Sirota et al.,[31] we observe the loss of crystallinity coinciding with the loss of transparency and the appearance of diffuse scattering when the samples' volume fraction became smaller than ~13%. While whenever we concentrate even more, we kept the angle-independent structural color in reflection and some transparency, and the iridescence disappeared, which is most likely due to the increasingly glassy structure. Here it should be noted that while Sirota et al. made similar SAXS observations, their samples were strongly scattering in the visible range.

The scattering intensities measured for the 30% FL+PS suspensions also showed many scattering orders but at first glance one has the impression that the sample is predominantly liquid, although on macroscopic inspection we observed iridescence under certain angles in reflection, which can only arise from local order. Indeed, when zooming into the 2D scattering image, we observe small but clear Bragg peaks (Fig. 3). After dividing out the form-factor (here using only the F(q) for the FL particles) we assigned the Miller indices (*hkl*) to the peaks in the resulting *S*(*q*), using again the fact that scattering vectors are related to the nearest-neighbor distance *a* via $q_i = (\frac{2\pi}{a\sqrt{2}})(h^2 + k^2 + l^2)^{1/2}$. Diffraction peaks in a single-species crystal with FCC symmetry occur only for combinations of all (*hkl*) being either even or odd, while for BCC to show the sum of the Miller indices must be even. However, in the case of our 1:1 mixture we sea peaks at all combinations of the (*hkl*). This means our similarly sized but on average differently charged FL and PS particles seem to form a CsCl lattice with simple cubic structure. It is obvious that the large width of the scattering peaks reflects the polydispersity and thus a small crystallite formation in our system. Nevertheless, it is interesting that these seemingly different particles do not segregate from each other. This can be rationalized by Hume-Rothery rules stating that if the difference in particle size is no more than 15%, then AB crystals will form. This is again in analogy to metallic crystals/glasses, which are, in this case, formed by two different species A and B.

**3.4 Electromagnetic Simulations.** Based on the SAXS measurement results, we carried out electromagnetic simulations to show that changing the lattice parameter of the FCC crystal sets the reflection color. For this purpose, we formed FCC lattices of dielectric spheres having a refractive index of 1.37 and set the background refractive index to 1.33 equating to the refractive index of water. Our simulations prove that despite the low refractive index contrast between the medium and the colloids, this system can still reflect the incoming radiation at a certain range of wavelengths, which is determined by the lattice constant (Figure 4). We also show that in principle the whole visible regime can be covered using our colloids, which further prove the feasibility of using our materials in color filters, reflectors, and optical coatings.

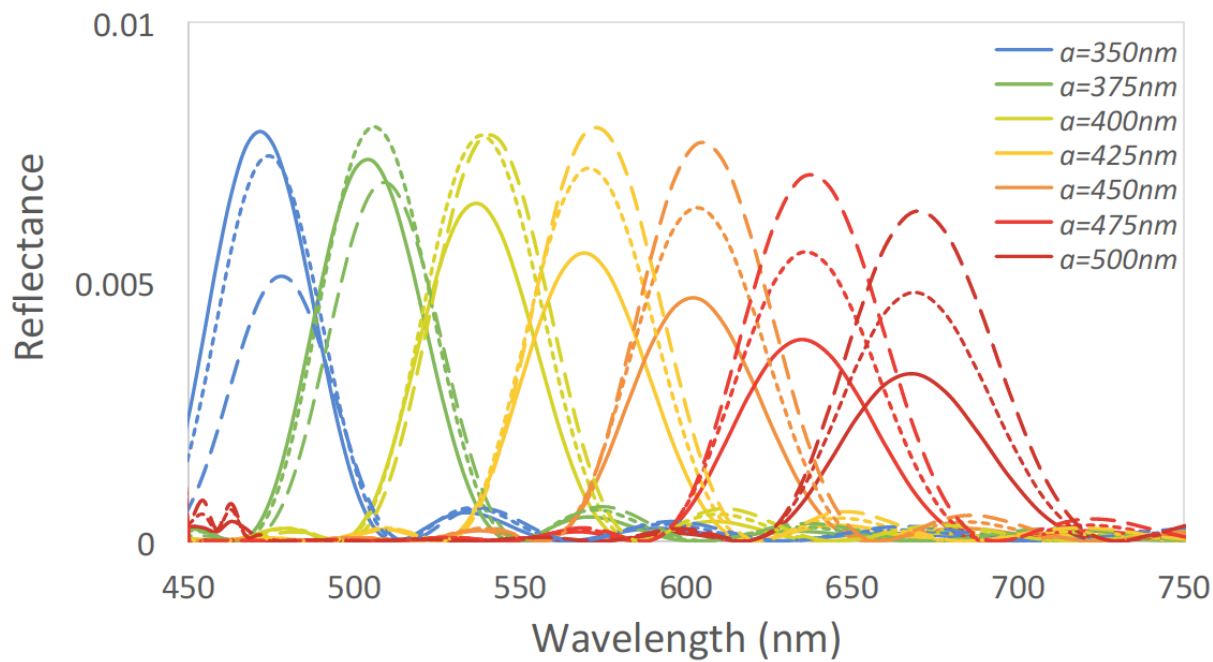

**Figure 4.** Simulated reflection spectra assuming an FCC crystal with different lattice constants *a* and particle diameters of 170 nm (continuous lines), 190 nm (short-dashed lines) and 210 nm (long-dashed lines).

It should be noted that we obtain iridescent colors also for FL particles that are only 150 nm or 250 nm in diameter, however with a small shift in color, with respect to a given volume fraction. However, the strength of the reflected colors seems to be most pronounced for the roughly 190 nm large ones. Also, suspensions with the smaller sized FL particles showed even high transparency. Finally, we were also able to prepare virtual '*inverted opal structures*' by increasing the refractive index of the continuous, aqueous phase to 1.4 by saturating it with sucrose. Again, partially transparent but iridescent structures could be observed.

**CONCLUSIONS**

In this article we have shown that it is possible to create transparent colloidal crystals that show structural color in reflection with aqueous suspensions of Fluorinated Latex particles having a very small refractive index difference. We studied the crystal structure of the system using SAXS and determined that these colloids self-organize into predominantly FCC lattice structures. We showed that their low refractive index helps to obtain a transparent colloidal photonic crystal, which has strong reflectance at a certain range of wavelengths within the visible regime. Using electromagnetic simulations, we confirmed our observations and showed that the reflection color has a strong dependence on the lattice parameter. Being transparent but also possessing a tunable reflection band, these colloidal photonic crystals hold great promise for low cost optical coatings, filters, and displays. Finally, we also show that 1:1 mixtures of similarly sized FL and PS particles with differently strong, negative surface charge rather form simple cubic structures, reminiscent of an CsCl crystal due to the difference in charging state and binary metallic crystals with AB lattice.

## ACKNOWLEDGEMENTS

TE is grateful to the Royal Society for supporting for the Newton International Fellowship. MZ thanks the Engineering and Physical Science Research Council (EPSRC) and Unilever for the CASE award RG748000. AC and EE acknowledge the ETN-COLLDENSE (H2020-MCSA-ITN-2014, Grant No. 642774) and the Winton Program for the Physics of Sustainability. TON acknowledges the doctoral training center (NanoDTC) program provided by the EPSRC.